\documentclass[aps,amssymb,12pt,floatfix]{revtex4}
\usepackage{subfigure}
\usepackage{graphicx}
\usepackage{rotating}

\begin{document}

\title{Probing protein-protein interactions by dynamic\\
force correlated spectroscopy (FCS)}

\author{V. Barsegov$^1$ and D. Thirumalai$^{1,2}$}
\thanks{Corresponding author phone: 301-405-4803; fax: 301-314-9404; thirum@glue.umd.edu}
\affiliation{$^1$Biophysics Program, Institute for Physical Science and Technology\\
$^2$Department of Chemistry and Biochemistry, University of Maryland, 
College Park, MD 20742}

\date{\today}

\begin{abstract}

We develop a formalism for single molecule dynamic force spectroscopy to map the 
energy landscape of protein-protein complex ($P_1$$P_2$). The joint distribution 
$P(\tau_1,\tau_2)$ of unbinding lifetimes $\tau_1$ and $\tau_2$ measurable in a 
compression-tension cycle, which accounts for the internal relaxation dynamics of 
the proteins under tension, shows that the histogram of $\tau_1$ is not Poissonian. 
The theory is applied to the forced unbinding of protein $P_1$, modeled as a wormlike 
chain, from $P_1$$P_2$. We propose a new class of experiments which can resolve the 
effect of internal protein dynamics on the unbinding lifetimes.


\end{abstract}

\maketitle

Many biological functions are mediated by interactions between biomolecules under 
mechanical stress. Protein-DNA interactions involve force-induced motion of proteins  \cite{1,2}. 
Similarly, specific protein-protein interaction in cell-protein complexes are important in 
molecular recognition \cite{2new}. Dynamic force spectroscopic techniques probe these interactions 
by forced unbinding of protein-protein complexes using forces in the $1pN$$-$$100pN$ range and can 
be used to map the complex energy landscape underlying protein-protein association 
\cite{1,2,3,4,5,11,12}. Atomic force microscopy (AFM) has been employed in the studies of 
protein-protein interactions involving immunoglobulins \cite{13}, molecular motors \cite{11,15}
and cell adhesion complexes \cite{2new,5,12}. 

In constant force-induced unbinding of single protein-protein complexes, the histograms
of unbinding lifetimes is fit using the Poisson distribution
\begin{equation}\label{1.3}
P_u(\tau;f_{ext})=k_1(f_{ext})\exp{[-k_1(f_{ext})\tau]}.
\end{equation}
The dependence of the unbinding rate constant $k_1$$=$$1/\tau_u$ ($\tau_u$ is 
the lifetime of the complex) on the external force $f_{ext}$ is given by the Bell model 
\cite{Bell}, $k_1(f_{ext})$$=$$k_{10}$$\exp{[f_{ext}\sigma/k_B T]}$. The parameter $\sigma$ 
is the maximum protein-protein bond extension before rupture, and $k_{10}$ is the 
force-free unbinding rate of the bound complex $P_1$$P_2$. Because the Poisson approximation 
ignores the intrinsic dynamics of proteins (i.e. conformational motions and rearrangements), this 
analysis can only be used when $\tau_u$ exceeds the timescale of internal protein motion, $\tau_R$. 
However, NMR studies of relaxation dynamics of proteins show that $\tau_R$ of single chain 
proteins ranges from nanoseconds to tens of miliseconds \cite{16new}. Lifetime measurements of 
a single P-selectin receptor with specific ligand PSGL-1 show that $\tau_u$ varies between 
miliseconds and few seconds depending on the magnitude of $f_{ext}$ \cite{5,2new}. Because the 
lifetimes of the protein-protein complex under force become comparable to $\tau_R$, the 
interpretation of the unbinding data is complicated by protein motion. Thus, Eq. (\ref{1.3}) 
cannot be used to describe experimental histograms of the lifetimes. To account for the competing 
timescales ($\tau_R$ and $\tau_u$) a theoretical framework that probes correlations between 
intrinsic relaxation and unbinding dynamics is needed to analyze experimental data.

In typical AFM experiments, the cantilever tip coated with protein $P_1$ is brought into 
contact with the surface-attached protein $P_2$, and allowed to interact for a time $\Delta t$ 
so that the complex $P_1$$P_2$ can form (compression cycle). The tip is then retracted to a 
prescribed distance which results in the complex feeling a constant force 
${\bf f}$$=$$f_{ext}$${\bf x}$ in the direction ${\bf x}$ perpendicular to the surface 
(tension cycle). The lifetime $\tau$ at which $P_1P_2$ bond breaks is recorded. However, if 
$\tau_u$$\sim$$\tau_R$, there is a finite time ($\sim$$\tau_R$) for propagation of the constant 
tension from the pulled terminus of $P_1$ to the binding interface of the $P_1$$P_2$ complex. 
Thus, the average time $\tau_u$ to break the $P_1$$P_2$ bond (assuming that cantilever spring 
constant is stiff compared with the non-covalent linkages that stabilize $P_1$ and $P_2$), is 
enhanced by $\tau_R$ resulting in the ``apparent'' lifetime $\tau$$\approx$$\tau_R$$+$$\tau_u$ 
of the complex. 

In this Letter we propose a novel theoretical methodology for describing forced unbinding  
which allows for accurate estimation of protein-protein interaction parameters. The approach 
is based on analyzing not only the distribution of single lifetimes $P(\tau)$ but also the 
joint distribution $P(\tau_1,\tau_2; \Delta t)$ of lifetimes $\tau_1$ and $\tau_2$ separated 
by compression time $\Delta t$. The distribution $P(\tau_1,\tau_2; \Delta t)$ is measurable by 
constructing the joint histogram of lifetimes using current experimental methods. Because in 
current AFM assays $\Delta t$ can be as short as microseconds \cite{17}, $\Delta t$ can be varied 
by changing the frequency of the compression cycle; $P(\tau_1,\tau_2; \Delta t)$ can be 
utilized to resolve $\tau_R$ which in turn can be used to obtain $\tau_u$, and free-energy 
landscape parameters $\sigma$ and $k_{10}$. The theory describes protein-protein complexes that 
obey $P_1$$+$$P_2$$\rightleftharpoons$$P_1$$P_2$, and can be extended to more elaborate 
kinetic and protein models.


{\it Basic concepts:} Typically, for specific protein-protein complexes the binding rate for 
$P_1$$+$$P_2$$\to$$P_1$$P_2$ is fast, and $\Delta t$ is controlled by the duration of the
compression cycle. Because of the conformational fluctuations of $P_1$, the binding 
interface experiences a restoring force $f(X,t)$ which tends to decrease the end-to-end distance 
$X(t)$. As $t$ increases, the unbinding force along the coordinate $X$ increases so that 
$f(X,t)$$\to$$f_{ext}$ as $t$$\to$$\infty$, and $X(t)$ approaches the equilibrium 
force-dependent value $\langle X(f_{ext})\rangle$. Due to the conformational dynamics of the 
proteins, the unbinding rate, $k_1(X,t)$$=$$k_{10}$$\exp{[\sigma f(X,t)/k_B T]}$, is a 
{\em stochastic variable} that depends on $X$ through $f(X)$. When application of $f_{ext}$ does 
not result in complete stretching of $P_1$ ($X$$=$$L$), the instantaneous value of force along 
the $P_1$$P_2$ bond is equal to the restoring force 
\begin{equation}\label{2.10}
f(X,\tau)=-k_B T {{1}\over {P(X,\tau)}}{{\partial P(X,\tau)}\over {\partial X}} 
\end{equation}
where the probability that $P_1$ has end-to-end distance $X$ at time $t$ is given by 
$P(X,t)$$=$${{1}\over {N(t)}}$$\int_0^L$$dX_0$$4\pi$$X_0^2$$G_0(X,t;X_0)$$\psi_{eq}(X_0)$ 
and $N(t)$ is a normalization constant. When $f_{ext}$ is large to fully stretch $P_1$, 
the force felt by $P_1$$P_2$ bond spikes up to $f_{ext}$ at $X$$=$$L$, i.e. 
$f$$=$$f(X,\tau)$$h(L-X)$$+$$f_{ext}$$h(X-L)$, where $h(X)$ is the Heavyside step function. In 
this Letter we consider $f_{ext}$ that does not exceed the unfolding force threshold. The 
unbinding time distribution is given by the convolution of unbinding kinetics and dynamics of 
$X$, i.e.,
\begin{equation}\label{1.6}
P(\tau,f_{ext})={{1}\over {N_1}}\int_0^L dX_1 4\pi X_1^2  \int_0^L dX_0 4\pi X_0^2  
P_u(X_1,\tau) G_{f_{ext}}(X_1,\tau; X_0)\psi_{eq}(X_0)
\end{equation}
where $N_1$ is a normalization constant. In Eqs. (\ref{2.10}) and (\ref{1.6}), 
$G_{0}(X_1,t; X_0)$ and $G_{f_{ext}}(X_1,t; X_0)$ are respectively the force-free and 
force-dependent conditional probability of $X$ at time $t$ and $\psi_{eq}(X)$
is the equilibrium distribution of $X$. The unbinding probability $P_u(X,t)$ depends on $X$ 
through $k_1$, i.e. $P_u(X,t)$$=$$k_1(X,t)$$\exp{[-k_1(X,t)t]}$. The above equation is a 
generalization of Eq. (\ref{1.3}) for force exerted on $P_1$$P_2$ bond that continuously 
evolves from zero to $f$$=$$f_{ext}$ over time $\tau_R$. In the limit 
$\tau_R$$\ll$$\tau_u$, $P(\tau,f_{ext})$ reduces to $P_u(\tau,f_{ext})$ given by Eq. (\ref{1.3}).

{\it A model application:} To illustrate the consequences of the stochastic nature of $k_1(X,t)$, 
which reflects the underlying dynamics of proteins, we assume that a thermally fluctuating 
worm-like chain ($P_1$) is in contact with the immobile $P_2$. Upon application of force $f_{ext}$,
extension of $P_1$ results in unbinding. In this example, the timescale for internal modes of $P_1$
are comparable to the unbinding lifetime. The Hamiltonian for $P_1$ is 
\begin{eqnarray}\label{1.13}
H & = & {{3k_B T}\over {2l_p}}\int_{-L/2}^{L/2} ds \left({{\partial {\bf r}(s,t)}
\over {\partial s}}\right)^2 +
{{3l_p k_B T}\over {8}}\int_{-L/2}^{L/2} ds\left({{\partial^2 {\bf r}(s,t)}
\over {\partial s^2}}\right)^2 
\\\nonumber
& + & {{3 k_b T}\over {4}}\left[ \left({{\partial {\bf r}(-L/2,t)}
\over {\partial s}}\right)^2 +
\left({{\partial {\bf r}(-L/2,t)}\over {\partial s}}\right)^2 \right]+ 
{\bf f}\int_{-L/2}^{L/2}ds \left({{\partial {\bf r}(s,t)}
\over {\partial s}}\right)
\end{eqnarray}
where $l_p$ is the protein persistence length and ${\bf r}(s,t)$ is the location of 
protein monomers $-L/2$$\le$$s$$\le$$L/2$ at time $t$. The end-to-end vector is 
${\bf X}(t)$$=$${\bf r}(L/2,t)$$-$${\bf r}(-L/2,t)$, where $L$ is the protein contour 
length. The statistics of $X$ can be represented by a large number of independent modes when 
$L/l_p$$\gg$$1$. Thus, it is reasonable to assume that $G_0(X,t;X_0)$ 
is a Gaussian. In the overdamped limit, when $f_{ext}$ exceeds the unfolding threshold force, 
stretching of $P_1$ is smooth and thus, preserves Gaussian statistics,
\begin{equation}\label{2.1}
G_0(X,t;X_0) = \left( {{3}\over {2 \pi \langle X^2\rangle}} \right)^{3/2} 
{{1}\over {(1-\phi^2(t))^{3/2}}} \exp{\left[-{{3(X-\phi(t)X_0)^2}
\over {2\langle X^2\rangle (1-\phi^2(t))}}\right]}
\end{equation}
specified by the mean value $\langle X(t)\rangle = \phi (t)X_0 $ and variance 
$\sigma^2$$=$$\langle X^2\rangle $$-$$\langle X\rangle^2$, where the correlation function
$\phi(t)$$=$${{\langle X(t)X(0)\rangle }/{\langle X^2 \rangle }}$. To construct $G_0(X,t;X_0,0)$ 
we compute $\langle X(t)X(0)\rangle$ and 
$\langle X^2\rangle$$=$$\lim_{t\to\infty}\langle X(t)X(0)\rangle$ with $f_{ext}$$=$$0$. 
By using Eq. (\ref{1.13}) and assuming that the dynamics of the worm-like chain in the overdamped 
random media is described by a stochastic force $\xi(s,t)$ with white noise statistics, 
$\langle \xi_{\alpha} (s,t) \rangle$$=$$0$ and $\langle \xi_{\alpha}(s,t)\xi_{\beta}(s',t')\rangle$
$=$$2$$\gamma$$k_B T$$\delta_{\alpha \beta}$$\delta(s-s')$$\delta(t-t')$, where $\alpha$$=$$x,y,z$ 
and $\gamma $ is the friction coefficient, we arrive at the Langevin equation:
\begin{equation}\label{2.6}
\gamma {{\partial}\over {\partial t}}{\bf r}(s,t) + \epsilon {{\partial^4}
\over {\partial s^4}}{\bf r}(s,t)
-2\nu {{\partial^2}\over {\partial s^2}}{\bf r}(s,t) = {\bf \xi}(s,t)
\end{equation}
where $\epsilon$$=$${{3l_p k_B T}/ {4}}$ and $\nu$$=$${{3 k_B T}/{2 l_p}}$. We solve Eq. (\ref{2.6})
for ${\bf r}(s,t)$ with boundary conditions 
$[2\nu {{\partial}\over {\partial s}}{\bf r}$$-$$\epsilon {{\partial^3}
\over {\partial s^3}}{\bf r}]_{\pm L/2}$$=$$0$, 
$[2\nu_0 {{\partial}\over {\partial s}}{\bf r}$$\pm$$\epsilon {{\partial^2}
\over {\partial s^2}}{\bf r}]_{\pm L/2}$$=$$0$, where $\nu_0$$=$$3k_B T/4$ to yield \cite{35}: 
\begin{equation}\label{2.8}
\langle X(t)X(0)\rangle_0 =12 k_B T \sum_{n=1}^{\infty}{{1}\over {z_n}}\psi^2_n(L/2) 
e^{-z_n t/\gamma}, \quad n=1,3,\ldots , 2q+1
\end{equation}
where the odd eigenfunctions are \cite{35}
\begin{equation}\label{A.4}
\psi_n(s)=\sqrt{{{c_n}/ {L}}}\left(  {{\alpha_n}\over {\cos{[\alpha_n L/2]}}} 
\sin{[\alpha_n s]} + {{\beta_n}\over {\cosh[\beta_n L/2]}} \sinh{[\beta_n s]}\right)
\end{equation}
with normalization constant $c_n$. The eigenvalues 
$z_n$$=$$\epsilon$$\alpha_n^4$$+$$2$$\nu$$\alpha^2$ 
and the constants $\alpha_n$, $\beta_n$ are obtained by solving 
$\alpha_n$$\sin{[{{\alpha_n L}\over {2}}]}$$\cosh{[{{\beta_n L}\over {2}}]}$$-$
$\beta_n^3$$\cos{[{{\alpha_n L}\over {2}}]}$$\sinh{[{{\beta_n L}\over {2}}]}$$-$
${{1}\over {l_p}}$$(\alpha_n^2+\beta_n^2)$$\cos{[{{\alpha_n L}\over {2}}]}$
$\cosh{[{{\beta_n L}\over {2}}]}$$=$$0$ and $\beta_n^2$$-$$\alpha_n^2$$=$${{1}\over {l_p^2}}$. 
In the limit, $L/l_p$$\to$$\infty$, we arrive at the Rouse chain model describing the 
stretching modes
$\psi_n^R$$=$$\sqrt{{{2}/ {L}}}$$\sin{\left( {{n\pi s}/ {L}} \right)}$ with eigenvalues 
$z_n^R$$=$${{3n^2\pi^2 k_B T}/{2l_p L^2}}$. To construct force-dependent propagator 
$G_{f_{ext}}(X,t;X_0)$, we integrate Eq. (\ref{2.6}) with $f_{ext}$${\bf x}$ added to 
${\bf \xi}(s,t)$ to obtain $\langle X^2 \rangle_{f_{ext}}$$=$$\langle X^2 \rangle_0$$+$
$f_{ext}^2$$\sum_{n=1}^{\infty}$$\psi_n^2(L/2)/z_n^2 $.
 
We computed $P(\tau,f_{ext})$ by integrating Eq. (\ref{1.6}) at room temperature. The parameters 
$L$, $l_p$ and $\gamma$$=$$k_B T/DL$ determine the timescale of protein motion 
$\tau_R$$\approx$$\text{max}\{ \gamma/z_n\}$. We set $k_{10}$$=$$0.1$$\mu s^{-1}$, 
$\sigma$$=$$1.0nm$, $L$$=$$80nm$, $l_p$$=$$0.4nm$ and $D$$=$$10^{-8}cm^2/s$. The 
largest eigenvalue $z_1/\gamma$$=$$0.2$$\mu s^{-1}$ determines the longest relaxation 
timescale $\tau_R$$\approx$$5\mu s$. In left panels of Figure 1 we compare $P(\tau,f_{ext})$ 
for WLC and Rouse model (Eq. (\ref{1.6})) with the Poisson approximation $P_u(\tau,f_{ext})$ 
(Eq. (\ref{1.3})) for $f_{ext}$$=$$1pN$, $3pN$ and $10pN$. At $f_{ext}$$=$$3pN$ and $10pN$, 
$P(\tau,f_{ext})$ for WLC model is in good agreement with $P(\tau,f_{ext})$ computed for the 
Rouse model. A slight overestimate in $P(\tau)$ at short $\tau $'s and lower $f_{ext}$$=$$1pN$ 
is due to faster relaxation of the Rouse modes. For $k_1$$\sim$$z_1/\gamma$, Poisson approximation 
$P_u(\tau)$ deviates noticeably from $P(\tau)$. Deviations grow as $f_{ext}$ 
is increased from $1 pN$ to $10 pN$; $P_u(\tau)$ overestimates $P(\tau)$ at shorter $\tau $ 
and underestimates $P(\tau)$ at longer $\tau $, predicting shorter lifetimes. 
Therefore, in cases when protein conformational relaxation and unbinding dynamics occur on
similar timescales the use of Poisson approximation leads to inaccurate estimates of $k_{10}$ and 
$\sigma $. In the right panels of Figure 1 we compare $P(\tau,f_{ext})$ for the WLC and Rouse modes
with Poisson approximation $P_u(\tau,f_{ext})$ for $z_1/\gamma$$=$$2$$\mu s^{-1}$$\gg$$k_{10}$. 
A tenfold increase in $z_1/\gamma$ corresponds to less overdamped environment with larger 
$D$$=$$10^{-7}cm^2/s$ (the other parameters are same as in left panels). Because, it now takes an 
order of magnitude shorter time to propagate $f_{ext}$ from the pulled end of $P_1$ to the 
$P_1$$P_2$ interface, Poisson distribution $P_u$ follows closely $P(\tau,f_{ext})$ at lower 
$f_{ext}$$=$$1pN$ and $3pN$. However, $P_u$ deviates from $P(\tau,f_{ext})$ at higher 
$f_{ext}$$=$$10pN$ due to rapid force-induced increase in the unbinding rate $k_1$. 
Thus, {\it even when propagation of tension is rapid there are substantial deviations from Poisson 
distribution of bond lifetimes at higher $f_{ext}$}.

 
A practical methodology that can be used in conjunction with experimental data to accurately 
estimate of $k_{10}$ and $\sigma $ is required. Dynamical signatures of protein motion can be 
assessed by computing the joint distribution $P(\tau_1,\tau_2; \Delta t)$ of consecutive unbinding 
times, $\tau_1$ and $\tau_2$, separated by compression time $\Delta t$,  
\begin{eqnarray}\label{1.7}  
P(\tau_1,\tau_2;\Delta t,f_{ext})
& = & {{1}\over {N_2}}\int_0^L dX_3 4\pi X_3^2 \int_0^L dX_2 4\pi X_2^2 
\int_0^L dX_1 4\pi X_1^2 \int_0^L 
dX_0 4\pi X_0^2 \\\nonumber
& \times & P_u(X_3,\tau_2) G_{f_{ext}}(X_3,\tau_2; X_2)P_b(X_2,\Delta t)
G_0(X_2,\Delta t;X_1)\\\nonumber
& \times & P_u(X_1,\tau_1)G_{f_{ext}}(X_1,\tau_1;X_0)\psi_{eq}(X_0)
\end{eqnarray}
where $P_b(t)$ is the binding probability for $P_1$$+$$P_2$$\to$$P_1$$P_2$ and $N_2$ is a 
normalization constant. In Eq. (\ref{1.7}), $G_0(X_2,t; X_1)$ is the force free propagator 
representing correlations of two interaction events decaying over $\tau_R$. When 
$\tau_R$$>$$\Delta t$, $P(\tau_1,\tau_2;\Delta t)$ is a sensitive measure of protein motion 
and thus, can be employed to estimate $\tau_R$. When $\tau_R$$\ll$$\Delta t$, unbinding events 
are independent, $\lim_{\Delta t \to \infty }G_0(X_2,\Delta t;X_1)$$\to$$\psi_{eq}(X_2)$, and 
hence, $P(\tau_1,\tau_2)$$\to$$P(\tau_1)$$P(\tau_2)$. 

We computed $P(\tau_1,\tau_2;\Delta t)$ for $\Delta t$$=$$1\mu s$$\ll$$\gamma/z_1$, 
$\Delta t$$=$$10 \mu s$$\sim$$\gamma/z_1$ and $\Delta t$$=$$500 \mu s$$\gg$$\gamma/z_1$ for 
$f_{ext}$$=$$3.0 pN$ and $k_{10}$$=$$0.1$$\mu s^{-1}$, $\sigma$$=$$1.0 nm$, $L$$=$$80nm$, 
$l_p$$=$$0.4nm$ and $z_1/\gamma$$=$$0.01$$\mu s^{-1}$ (Fig. 2). We assumed that 
protein binding ($P_1$$+$$P_2$$\to$$P_1$$P_2$) is independent of the dynamics of $X$, i.e. 
once $P_1$ reached the vicinity of binding interface of $P_2$ it binds, and set 
$P_b(X,\Delta t)$$=$$P_b$$=$$1$ in Eq. (\ref{1.7}). A short $\Delta t$$=$$1\mu s$ and $10\mu s$ 
peak in $P(\tau_1,\tau_2)$ (top and middle panels) is washed out at longer $\Delta t$$=$$500\mu s$ 
(bottom). Striking asymmetry of the contour plots at short $\Delta t$ is due to the dependence of 
shorter $\tau_2$-events on longer $\tau_1$-events. During the first interaction the constant force 
felt by $P_1$$P_2$ bond is ramped up from $f$$=$$0$ to $f$$=$$f_{ext}$ following the restoring 
force $f(X,t)$ thus, prolonging $\tau_1$. When $\Delta t$$\ll$$\tau_R$$\sim$$\gamma/z_1$, the 
next binding event takes place (at $t$$=$$\Delta t$ after the first unbinding) when $P_1$ is 
partially or fully stretched. As a result, the binding interface experiences non-vanishing 
restoring force from the beginning of the second interaction event and $\tau_2$$<$$\tau_1$. 
Contour plots of $P(\tau_1,\tau_2)$ become more symmetric as $\Delta t$ is increased to $10\mu s$ 
which implies growing statistical independence of unbinding events. At 
$\Delta t$$=$$500 \mu s$$\gg$$\tau_R$, $P(\tau_1,\tau_2)$ is symmetric density, which results in 
factorization $P(\tau_1,\tau_2)$$=$$P(\tau_1)$$P(\tau_2)$. Thus, to obtain {\em statistically 
meaningful distributions of uncorrelated unbinding times}, unbinding events must be separated by 
much longer $\Delta t$ compared to $\tau_R$ whose a priori determination is difficult.


{\it Application to Experiments:} Using 
$D(\tau_1,\tau_2;\Delta t )$$=$$P(\tau_1,\tau_2;\Delta t)$$-$$P(\tau_1)$$P(\tau_2)$,
correlations between $\tau_1$'s and $\tau_2$'s can be probed in AFM experiments. If $D$$\neq$$0$, 
the unbinding events are influenced by conformational fluctuations of the protein. For the model
parameters in Fig. 2 we show in Fig. 3 $D(\tau_1=\tau_2;\Delta t )$ for $\Delta t$$=$$1\mu s$, 
$10\mu s$ and $500\mu s$. The peak of $D(\tau,\Delta t)/D(\tau,0)$, 
which signifies the amplitude of correlations between the unbinding events, decays to zero
as $\Delta t$ is increased from $1\mu s$$\ll$$\gamma/z_1$ to $500\mu s$$\gg$$\gamma/z_1$.
An accurate statistical analysis of unbinding lifetimes can be made using the following steps. 
From the unbinding time histogram $P(\tau,f_{ext})$ and the ``apparent'' mean lifetime $\tau_{app}$ 
the joint histogram $P(\tau_1,\tau_2; \Delta t)$ for $\Delta t$$\ll$$\tau_{app}$ is computed. 
The difference $D(\tau_1,\tau_2;\Delta t)$ is evaluated using $P(\tau,f_{ext})$ and {\it the 
experimentally determined $P(\tau_1,\tau_2;\Delta t,f_{ext})$}. If $D$$\approx$$0$, the 
unbinding events are uncorrelated, and $k_1$ can be estimated by fitting Eq. (\ref{1.3}) to 
$P(\tau,f_{ext})$. If $D$$>$$0$, the unbinding and protein motions are correlated. In this case 
the lifetime measurements must be repeated for longer $\Delta t$. Using the new data, new 
distributions $P(\tau_1)$, $P(\tau_1,\tau_2; \Delta t)$ and $D(\tau_1,\tau_2;\Delta t)$ can be 
calculated. The process is iterated until the requirement $D$$\approx$$0$ is satisfied for the 
compression cycle of duration, say, $\Delta t^*$. The protein relaxation time $\tau_R$ is the 
minimum value of $\Delta t$$=$$\Delta t^*$ at which $D$$\approx$$0$. Uncorrelated lifetimes 
collected for $\Delta t$$\gg$$\tau_R$$\approx$$\Delta t^*$ can then be binned to obtain the final 
histogram $P(\tau)$. If $\tau_R$$\ll$$\tau_{app}$$=$$\tau_R$$+$$\tau_u$ then 
$\tau_{app}$$\approx$$\tau_u$, and $k_{10}$ and $\sigma$ can be estimated by fitting 
Eq. (\ref{1.3}) to $P(\tau,f_{ext})$. However, if $\tau_R$$\sim$$\tau_{app}$, 
$P(\tau,f_{ext})$ must be analyzed using Eq. (\ref{1.6}) for given $f_{ext}$, $L$, 
$\gamma$$=$$k_BT/DL$, and estimated $\tau_R$. Thus, the theory presented here suggests a novel 
dynamic correlated force spectroscopy in which measurements of $P(\tau_1,\tau_2;\Delta t)$ for 
varying $\Delta t$ can be used to account for the influence of internal protein dynamics on forced 
unbinding of protein-protein complexes. 

This work was supported by National Science Foundation Grant NSFCHE-05-14056.


\newpage

\section*{\bf FIGURE CAPTIONS}

{\bf Figure 1}
The distribution of unbinding times $P(\tau,f_{ext})$ for WLC (solid) and Rouse model 
(dash-dotted lines) of protein and Poisson approximation $P_u(\tau,f_{ext})$ (dashed lines) 
for $f_{ext}=1pN$, $3pN$ and $10pN$ computed for $k_1$$\sim$$z_1/\gamma$ (left) and 
$k_1$$\ll$$z_1/\gamma$ (right panels).

\bigskip

{\bf Figure 2}
The joint distribution $P(\tau_1,\tau_2; \Delta t,f_{ext})$ of lifetimes $\tau_1$ and 
$\tau_2$ separated by $\Delta t$$=$$1\mu s$ (top), $10\mu s$ (middle) and $500 \mu s$ 
(bottom panels) for $f_{ext}$$=$$3pN$. The contour plots of $P(\tau_1,\tau_2;\Delta t,f_{ext})$ 
are shown on the right.

\bigskip

{\bf Figure 3}
Normalized correlation amplitude $D(\tau,\Delta t)/D(\tau,0)$ of equal lifetimes 
$\tau$$=$$\tau_1$$=$$\tau_2$ separated by $\Delta t$$=$$1\mu s$ (solid), $10\mu s$ (dash-dotted) 
and $500 \mu s$ (dashed lines) for $f_{ext}$$=$$3pN$. 

\newpage

\begin{figure}
\includegraphics[width=7.00in]{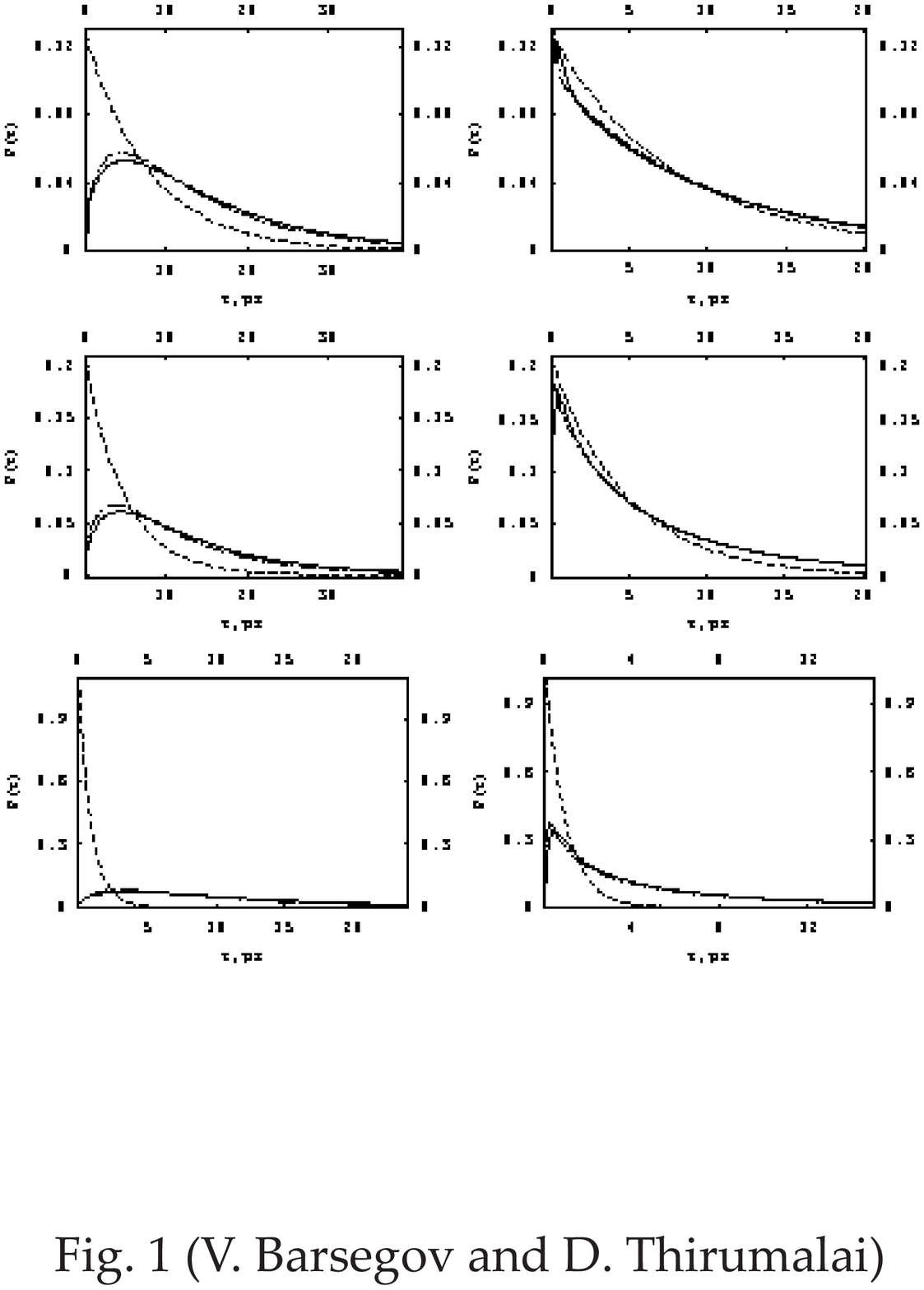}
\end{figure}

\newpage

\begin{figure}
\includegraphics[width=7.00in]{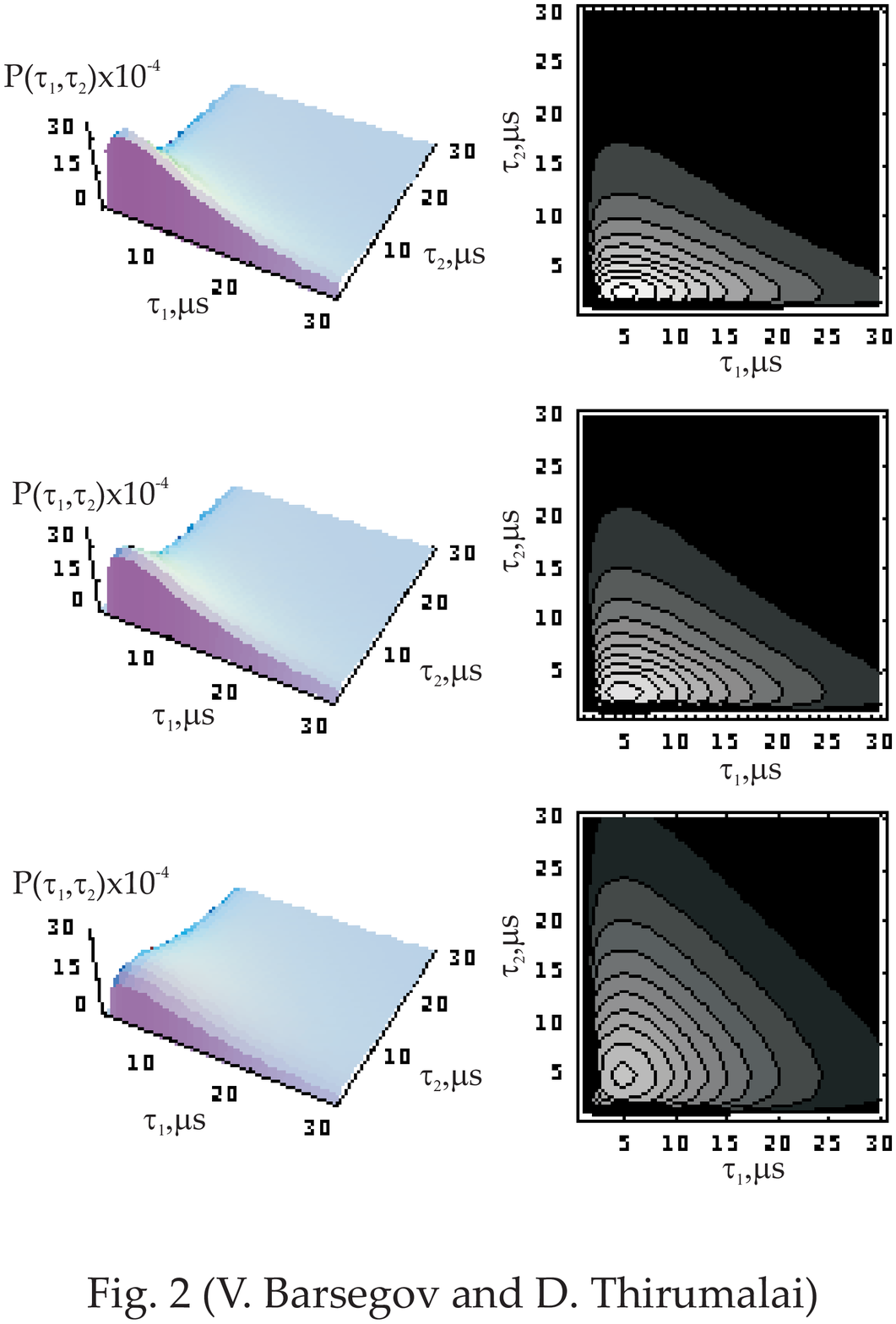}
\end{figure}

\newpage

\begin{figure}
\includegraphics[width=7.00in]{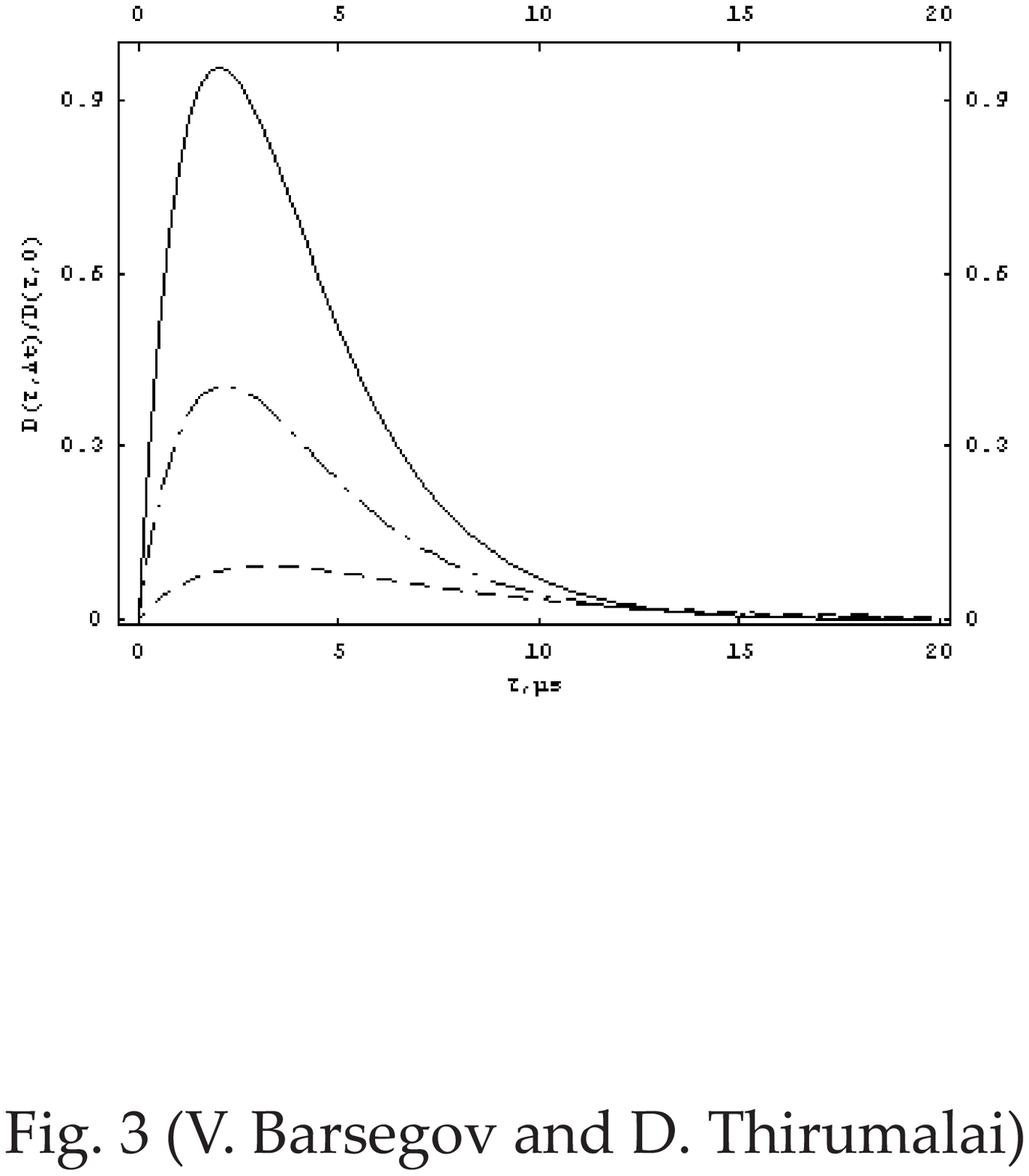}
\end{figure}

\end{document}